\documentclass[%
reprint,
preprintnumbers,
nofootinbib,
amsmath,amssymb,longbibliography,
aps
prd
]{revtex4-1}
\pdfoutput=1
\usepackage{graphicx}
\usepackage[utf8]{inputenc}
\usepackage{flushend}
\usepackage{dcolumn}
\usepackage{bm}
\usepackage{soul}
\usepackage{balance}
\usepackage[normalem]{ulem}
\usepackage[colorlinks = true,
            linkcolor = blue,
            urlcolor  = blue,
            citecolor = blue,
            anchorcolor = blue]{hyperref}
\usepackage{verbatim,subfigure}
\usepackage{color,ulem}
\usepackage[english]{babel}
\usepackage{MnSymbol,wasysym}
\usepackage[utf8]{inputenc}
\input Starburst.fd
\newcommand*\initfamily{\usefont{U}{Starburst}{xl}{n}}\initfamily 

\newcommand{\beq}{\begin{eqnarray}}
\newcommand{\eeq}{\end{eqnarray}}
\usepackage{amsmath}
\usepackage{tikz}
\usetikzlibrary{decorations.pathmorphing}
\usetikzlibrary{shapes.misc}
\tikzset{cross/.style={cross out, draw=black, minimum size=8*(#1-\pgflinewidth), inner sep=0pt, outer sep=0pt},
cross/.default={1pt}}
\usetikzlibrary{patterns,math}
\definecolor{applegreen}{rgb}{0.55, 0.71, 0.0}
\usepackage{contour}
\usepackage{xparse}
\ExplSyntaxOn
\NewDocumentCommand{\HS}{m}
 {
  \seq_set_split:Nnn \l_tmpa_seq { ~ } { #1 }
  \seq_map_inline:Nn \l_tmpa_seq { \contour{green}{##1} ~ } \unskip
 }
\ExplSyntaxOff

\definecolor{darkviolet}{rgb}{0.58, 0.0, 0.83}
\definecolor{mygreen}{rgb}{0.0, 0.5, 0.0}

\begin{document}
\preprint{\texttt{IFT-UAM/CSIC-24-107}}

%
\title{Krylov complexity as an order parameter for quantum chaotic-integrable transitions}

\author{Matteo Baggioli,$^{1,2,3}$ Kyoung-Bum Huh,$^{2}$ Hyun-Sik Jeong,$^{4}$ Keun-Young Kim$^{5,6}$ and Juan F. Pedraza$^{4}$}

\affiliation{\vspace{4pt}$^{1}$School of Physics and Astronomy, Shanghai Jiao Tong University, Shanghai 200240, China}
\affiliation{$^{2}$Wilczek Quantum Center, School of Physics and Astronomy, Shanghai Jiao Tong University, Shanghai 200240, China}
\affiliation{$^{3}$Shanghai Research Center for Quantum Sciences, Shanghai 201315, China}
\affiliation{$^{4}$Instituto de F\'isica Te\'orica UAM/CSIC, Calle Nicol\'as Cabrera 13-15, 28049 Madrid, Spain}
\affiliation{$^{5}$Department of Physics and Photon Science, Gwangju Institute of Science and Technology, Gwangju 61005, Korea}
\affiliation{$^{6}$Research Center for Photon Science Technology, Gwangju Institute of Science and Technology, Gwangju 61005, Korea}

\begin{abstract}
Krylov complexity has recently emerged as a new paradigm to characterize quantum chaos in many-body systems. However, which features of Krylov complexity are prerogative of quantum chaotic systems and how they relate to more standard probes, such as spectral statistics or out-of-time-order correlators (OTOCs), remain open questions. Recent insights have revealed that in quantum chaotic systems Krylov state complexity exhibits a distinct peak during time evolution before settling into a well-understood late-time plateau. In this work, we propose that this Krylov complexity peak (KCP) is a hallmark of quantum chaotic systems and suggest that its height could serve as an `order parameter' for quantum chaos. We demonstrate that the KCP effectively identifies chaotic-integrable transitions in two representative quantum mechanical models at both infinite and finite temperature: the mass-deformed Sachdev-Ye-Kitaev model and the sparse Sachdev-Ye-Kitaev model. Our findings align with established results from spectral statistics and OTOCs, while introducing an operator-independent diagnostic for quantum chaos, offering more `universal' insights and a deeper understanding of the general properties of quantum chaotic systems.\\
\end{abstract}

\maketitle
%
\section{Introduction}
Chaos is a widespread phenomenon in nature. While substantial progress has been made in understanding classical chaos \cite{hilborn2000chaos}, its definition and characterization in the quantum realm, particularly in many-body systems, remain significantly less understood.

Traditionally, quantum chaos has been linked to the Bohigas-Giannoni-Schmit (BGS) conjecture~\cite{Bohigas:1983er,Bohigas:1984aa,Guhr:1997ve,Bohigas}, asserting that the energy spectra of quantum systems with chaotic classical counterparts match the statistical predictions of random matrix theory (RMT). Specifically, quantum chaotic systems are expected to display RMT features such as level repulsion and spectral rigidity~\cite{Bohigas:1983er,Berry1985-mx,Muller:2004nb}, which are therefore accepted as fingerprints of late-time quantum chaos in many-body systems.

Conversely, in quantum chaotic systems with many degrees of freedom, ranging from the SYK model to black holes and other large-$N$ systems, at early times, the time evolution of specific out-of-time-order correlators (OTOCs) exhibits a phase of exponential growth \cite{larkin1969quasiclassical,berman1978condition}, governed by a non-zero Lyapunov exponent $\lambda_L \leq 2\pi k_B T/\hbar$ \cite{Maldacena_2016}. This behavior serves as an additional indicator of quantum chaos at complementary time scales.

Modern explorations of quantum chaos have prominently featured both level statistics and OTOCs, bolstered by intriguing links between many-body chaos and quantum gravitational systems~\cite{Shenker:2013pqa,Shenker:2014cwa,PhysRevD.94.126010,Cotler:2016fpe,deBoer:2017xdk,Stanford:2019vob}.
In this context, Krylov complexity~\cite{Parker:2018yvk,Balasubramanian:2022tpr} has recently emerged as a new valuable tool for characterizing quantum chaos, providing an alternative diagnostic beyond traditional analyses. It has been employed in RMT \cite{Balasubramanian:2022tpr,Balasubramanian:2023kwd,Tang:2023ocr,Caputa:2024vrn,Bhattacharjee:2024yxj} and many other quantum chaotic systems, including quantum billiards~\cite{Hashimoto:2023swv,Camargo:2023eev,Balasubramanian:2024ghv}, spin chains~\cite{Rabinovici:2021qqt,Scialchi:2023bmw,Gill:2023umm,Bhattacharya:2023xjx,Camargo:2024deu,Scialchi:2024zvq}, and various flavors of the Sachdev–Ye–Kitaev (SYK) model~\cite{Rabinovici:2020ryf,Bhattacharjee:2022ave,Hornedal:2022pkc,Erdmenger:2023wjg,Chapman:2024pdw}. Additionally, Krylov complexity has been discussed in several other contexts including topological and quantum phase transitions~\cite{Caputa:2022eye,Afrasiar:2022efk,Caputa:2022yju,Pal:2023yik}, quantum batteries \cite{Kim:2021okd}, bosonic systems describing ultra-cold atoms \cite{Bhattacharyya:2023dhp}, saddle-dominated scrambling~\cite{Bhattacharjee:2022vlt,Huh:2023jxt}, and open quantum systems~\cite{Bhattacharya:2022gbz,Bhattacharjee:2022lzy,Mohan:2023btr,Bhattacharya:2023zqt,Bhattacharjee:2023uwx,Carolan:2024wov} among others -- see \cite{Nandy:2024htc} for a comprehensive review. Two forms of Krylov complexity have been proposed in the literature: the original type, which addresses operator growth \cite{Parker:2018yvk}, and a newer version, which evaluates the spread of a time-evolving quantum state within a specific subspace of the Hilbert space \cite{Balasubramanian:2022tpr}. The latter will be the focus of this manuscript. 

For time-evolved thermofield double (TFD) states in RMT (see \cite{Camargo:2024deu} for a discussion on state-dependence of Krylov complexity), Krylov state complexity exhibits four distinct phases: an initial linear ramp, a peak, a subsequent decline, and a plateau. According to \cite{Balasubramanian:2022tpr}, the peak overshooting the plateau followed by a decline appears to be a universal characteristic of quantum chaotic many-body systems and is therefore expected to be absent in integrable systems. In this context, the importance of the TFD state is motivated not only by its role in its connection between Krylov complexity and other chaos probes, such as the spectral form factor (discussed in detail later in the text), but also for its potential relevance in holography, where it serves as the holographic dual of a two-sided black hole.

Building on these observations, this paper provides further evidence that the Krylov complexity peak (KCP) is a defining feature of quantum chaos, 
potentially serving as an `order parameter' for quantum chaotic phases.  Simply put, we propose that the KCP vanishes in integrable systems and that its height exhibits critical dynamics, capable of diagnosing quantum chaotic to integrable transitions at both infinite and finite temperature.

To provide evidence for our proposal, we turn to the SYK model and its variants, which serve as useful toy models. Specifically, we focus on two representative examples: the mass-deformed SYK~\cite{Song_2017,Eberlein_2017,Garcia-Garcia:2017bkg} and the sparse SYK~\cite{Xu:2020shn,Garc_a_Garc_a_2021} models, which have been extensively studied for their relevance to quantum chaotic to integrable transitions, and have been thoroughly characterized through both level statistics and properties of the OTOCs~\cite{Song_2017,Eberlein_2017,Garcia-Garcia:2017bkg,Xu:2020shn,Garc_a_Garc_a_2021,Nosaka2018,Kim:2020mho,Garcia-Garcia:2020dzm,Lunkin:2020tbq,Nandy:2022hcm,Menzler:2024atb,C_ceres_2022,Orman:2024mpw,Caceres:2023yoj,Garcia-Garcia:2023jlu}. 

This paper is structured as follows: Section \ref{sec:2} introduces the SYK models under consideration. Section \ref{sec:3} provides an overview of Krylov complexity and the spectral form factor. In Section \ref{sec:4}, we show that Krylov complexity effectively identifies chaotic-integrable transitions in these models, in line with the appearance and disappearance of the ramp in the spectral form factor. Finally, Section \ref{sec:5} presents our main conclusions

\section{Computational Models\label{sec:2}}
\subsection{Mass-deformed SYK model}
As a first example, we consider the mass-deformed SYK model~\cite{Garcia-Garcia:2017bkg}, which involves $N$ fermions in $0+1$ dimensions. This model extends the original SYK model~\cite{RevModPhys.94.035004} by including an additional quadratic term, known as the random mass term, alongside the random quartic interactions. The Hamiltonian of the model is given by 
\begin{align}\label{GSYK}
    H = \frac{1}{4!}\sum^N_{i,j,k,l=1}\, J_{ijkl}\, \chi_i\, \chi_j\, \chi_k\, \chi_l\, + \frac{i}{2!}\, \sum^N_{i,j=1}\,\kappa_{ij}\, \chi_i\,\chi_j \,.
\end{align}
Here $\chi_i$ are Majorana fermions satisfying $\{\chi_i,\, \chi_j\} = \delta_{ij}$, residing in a Hilbert space of dimension $2^{\frac{N}{2}}$. The coupling constants $J_{ijkl}$ and $\kappa_{ij}$ are Gaussian-distributed random variables with zero mean. Their standard deviations are $\sqrt{6}J/N^{3/2}$ and $\kappa/\sqrt{N}$, respectively.

In the absence of a mass deformation, the SYK model is maximally chaotic and saturates the Maldacena-Shenker-Stanford bound on quantum chaos~\cite{Maldacena_2016}. Conversely, the purely quadratic Hamiltonian corresponds to an integrable system, inherently lacking any chaotic hallmark. As the variance of the random mass deformation $\kappa$ increases, the model transitions from chaotic to integrable, effectively detected through level statistics~\cite{Garcia-Garcia:2017bkg}. More precisely, the analysis of level spacing distribution and $r$-parameter statistics yield a critical value $\kappa_c\approx66$ for $\beta=0$. This transition is further confirmed by the out-of-time-ordered correlator (OTOC) and the value of the Lyapunov exponent, though there are open discussions on this point~\cite{PhysRevLett.126.109101,Garcia-Garcia:2020dzm}. Moreover, it has been analytically proven~\cite{PhysRevResearch.3.013023} that for $\kappa>\kappa_c$, all states are many-body localized, and spectral correlations are well described by Poisson statistics, as expected for an integrable system.

%
\subsection{Sparse SYK model}
As a second example, we consider the sparse SYK model~\cite{Xu:2020shn,
Garc_a_Garc_a_2021}. In this case, the Hamiltonian is
\begin{align}\label{SSYK}
    H = \frac{1}{4!}\sum^N_{i,j,k,l=1}\, x_{ijkl}\, J_{ijkl}\, \chi_i\, \chi_j\, \chi_k\, \chi_l\,,
\end{align}
where $\chi_i$ are Majorana fermions satisfying $\{\chi_i,\, \chi_j\} = \delta_{ij}$. The coupling constants $J_{ijkl}$ are Gaussian-distributed random variables with zero mean and standard deviation
\begin{align}
    \sigma=\sqrt{\frac{6 J^2}{p N^{3}}} \,.
\end{align}
Furthermore, $x_{ijkl}$ equals 1 with probability $p$ and 0 with probability $1-p$. The parameter $p$ determines the number of non-zero terms in the Hamiltonian, $kN$, given by
\begin{align}\label{k_order}
    kN=p\,\binom{N}{4}\,,
\end{align}
which controls the amount sparseness of the model.
For $p=1$, the model reduces to the standard SYK model. As $p$ (or equivalently $k$) decreases, the model transitions towards an integrable regime, as evidenced by level statistics analysis~\cite{C_ceres_2022,Orman:2024mpw,Garc_a_Garc_a_2021}, which identifies a critical value of $k_c\approx 1$ for this transition. This transition is further reflected in the behavior of the Lyapunov exponent, indicated by a significant reduction in the exponential growth of OTOCs~\cite{Caceres:2023yoj}. In a parallel analysis it was shown that the emergence of gravitational physics at low temperatures ---specifically the onset of Schwarzian dynamics--- requires $k$ to lie within the range of 1/4 to 4~\cite{Xu:2020shn}.

The Krylov complexity of states has been examined in the sparse SYK model, as discussed in \cite{Jha:2024nbl}, though from a different perspective. In contrast to their approach, we will use the TFD state as our initial condition, aligning with the original conjecture for Krylov complexity~\cite{Balasubramanian:2022tpr}. Additionally, we will extend the analysis to investigate the temperature dependence of this complexity.

\vspace{6pt}
Overall, these two variants of the SYK model serve as ideal playgrounds for exploring the features of Krylov complexity that are unique to quantum chaotic systems and examining how these features evolve as the system transitions to an integrable phase. For both models, we will take $J=1$ for our computations.

%
\section{Krylov complexity and \\ spectral form factor\label{sec:3}}
The calculation of Krylov complexity involves constructing the Krylov basis $\{|K_n \rangle\}$, achieved through the so-called Lanczos algorithm~~\cite{Lanczos:1950zz,RecursionBook}. This procedure yields two sets of Lanczos coefficients, $\{a_n,\,b_n\}$, which encode all information regarding the system's dynamics. These coefficients correspond to the tridiagonal elements of the Hamiltonian when expressed in the Krylov basis:
\begin{align}\label{}
    H|K_n \rangle = a_n | K_n \rangle + b_{n+1} | K_{n+1} \rangle + b_n | K_{n-1} \rangle \,.
\end{align}

Given the Lanczos coefficients, the Krylov wave functions $\psi_n(t)$ satisfy the iterative differential equation:
\begin{align}\label{DES}
    i \, \partial_t \psi_n(t) = a_n \psi_n(t) + b_{n+1} \psi_{n+1}(t) + b_n \psi_{n-1}(t) \,,
\end{align}
which represents the Schrödinger equation within the Krylov space governed by the Hamiltonian $H$, such that the time-evolved state is given by $|\psi(t) \rangle = \sum_n \psi_n(t) | K_n \rangle$.

Finally, Krylov complexity is defined as 
\begin{align}\label{eq:Krylov complexity}
    C(t) := \sum_{n} n \, |\psi_n(t)|^2 \,.
\end{align}
It measures the average depth of a time-evolving state in the Krylov basis, reflecting the spread of the wave function in this basis.
     
As initial state, we consider the TFD state, 
\begin{align}\label{eq:TFD state}
    |\psi(0) \rangle = \frac{1}{\sqrt{Z(\beta)}}\sum_n e^{-\frac{\beta E_n}{2}} | n\rangle\otimes\vert\,n \rangle\,,
\end{align}
with $|n\rangle$ and $E_n$ indicating respectively the eigenstates and the eigenvalues of the Hamiltonian $H$. This TFD state is built from the tensor product of two copies of the original Hilbert space. Here, $Z(\beta)= \sum_n e^{-\beta E_n}$ is the partition function at inverse temperature $\beta$.\footnote{In the Hessenberg form using Householder reflections, the standard initial state is chosen as $(1, 0, 0, \cdots)^{\text{T}}$. Therefore, a basis transformation is required to align the given initial vector with this standard state. For more details, see \cite{Balasubramanian:2022tpr}.} Recall that the TFD state evolves under ($H_L+H_R$)/2, where $H_{L}$ and $H_{R}$ act independently on the left and right copies of the Hamiltonian, respectively.

The spectral form factor (SFF) is another valuable tool to probe the dynamics of quantum chaos, which may be linked to level statistics. Specifically, for quantum systems with discrete energy levels $\{E_n\}$, the SSF is defined via the analytically continued partition function~\cite{Guhr:1997ve,Brezin:1997rze},
\begin{align}\label{eq:SFF}
    \textrm{SFF}(t) :=&\, \frac{\vert Z(\beta+it)\vert^2}{\vert Z(\beta)\vert^2}\cr =&\, \frac{1}{Z(\beta)^2}\sum_{m,n}e^{-\beta(E_{m}+E_{n})}e^{i(E_{m}-E_{n})t} \,.
\end{align}
Both in RMT and in the SYK model, $\textrm{SFF}(t)$ displays a characteristic slope-dip-ramp-plateau behavior.

Given the definitions provided in Eq.\eqref{eq:Krylov complexity} and Eq.\eqref{eq:SFF}, it is natural to assume that a direct link between Krylov complexity and the SFF might exist. In fact, considering the Krylov complexity of the TFD state, the SFF can be interpreted as the survival probability of the time-evolved TFD state \cite{delCampo:2017bzr,Caputa:2024vrn}: $\textrm{SFF}(t) = |\psi_0(t)|^2$, where $\psi_{n=0}(t)$ is determined by solving Eq.\eqref{DES}. Additionally, for the TFD state with $\beta=0$, which is a maximally-entangled state, the late-time behavior of the Krylov complexity obeys the following identity \cite{Cotler:2016fpe,Rabinovici:2020ryf,Rabinovici:2022beu,Erdmenger:2023wjg}
\begin{align}\label{eq:saturation}
 \lim_{T\rightarrow \infty } \frac{1}{T}\int_{0}^{T}\text{SFF}(t) = \frac{1}{1+2C(t\rightarrow\infty)} \,,
\end{align}
where $C(t=\infty) = {(d-1)}/{2}$, with $d$ being the system size related to $N$ as $d=2^{N/2 - 1}$.\footnote{In the case of the TFD state with finite $\beta$, i.e., when the system is no longer in a maximally entangled state, a simple relation as in Eq.~\eqref{eq:saturation} may no longer hold. For further details on this point, see Ref. \cite{Erdmenger:2023wjg}.}
This identity provides a non-local (in time) constraint relating SFF$(t)$ to $C(t)$, reminiscent of a sum rule.

%
\section{Results\label{sec:4}}  
\subsection{Mass-deformed SYK model}
We calculate the time-dependent Krylov complexity, as defined in Eq.\eqref{eq:Krylov complexity}, for the mass-deformed SYK model as a function of the mass parameter $\kappa$ and the inverse temperature $\beta$. We perform our computations with a system size of $N=26$ and express all dimensionful quantities in units of the coupling $J$, following Ref.\cite{Garcia-Garcia:2017bkg}. As noted in \cite{Balasubramanian:2022tpr}, $N=26$ is sufficiently large to ensure convergence of the numerical results. Therefore, finite-size effects on Krylov complexity are minimal for this choice of $N$. For further details on the numerical methods, we refer the reader to the Appendix \ref{sec:App}.
\begin{figure}[t!]
 \centering
     {\includegraphics[width=\linewidth,trim={0 0.1cm 0 0},clip]{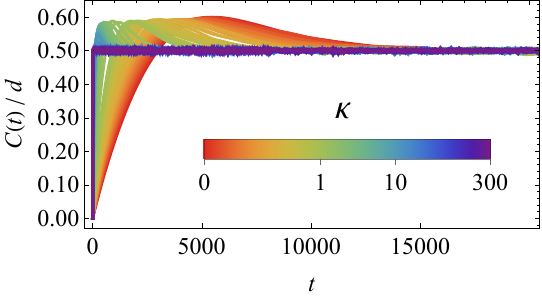} }
     {\includegraphics[width=\linewidth,trim={0 0.1cm 0 0},clip]{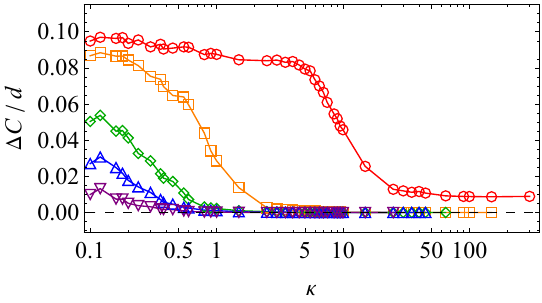} }\vspace{-0.3cm}
\caption{\textbf{Top:} Normalized Krylov complexity $C(t)/d$ with the TFD as initial state for various values of the parameter $\kappa = 0\, (\color{red}\textbf{red} \color{black})$ to $300\, (\color{darkviolet}\textbf{purple} \color{black})$ for $N=26$ and $\beta=0$. \textbf{Bottom:} Normalized difference between the peak value $C(t=t_{\text{peak}})$ and the late time value $C(t\rightarrow\infty)$, as a function of $\kappa$ for $\beta=0,\,1,\,3,\,5,\,10$ (\color{red}\textbf{red}\color{black}, \color{orange}\textbf{orange}\color{black}, \color{mygreen}\textbf{green}\color{black}, \color{blue}\textbf{blue}\color{black}, \color{darkviolet}\textbf{purple}\color{black}).} \label{Fig:Spread}
\end{figure}

In the top panel of Fig.~\ref{Fig:Spread}, we show the normalized Krylov complexity $C(t)/d$ as a function of time for various values of $\kappa \in [0,300]$, with colors ranging from \color{red}\textbf{red} \color{black} to \color{darkviolet}\textbf{purple} \color{black} at infinite temperature ($\beta=0$).  For low values of $\kappa$, Krylov complexity distinctly exhibits the four hallmark stages of chaotic dynamics: it first undergoes a linear ramp up to a peak at $t = t_{\text{peak}}$, followed by a decline that levels off into a constant plateau. The value of $C(t)$ at the plateau as $t \rightarrow \infty$ is independent of the parameter $\kappa$, only depending on the system size, with $C(t\rightarrow \infty)/d \approx 1/2$. Furthermore, as $\kappa$ increases, we note the KCP occurs at progressively earlier times and eventually vanishes when $\kappa$ becomes sufficiently large.

These observations indicate that the peak in Krylov complexity could be used as a clear indicator for quantum chaos, disappearing when the system becomes integrable, \textit{i.e.}, for large $\kappa$. To formalize this idea, we propose an `order parameter' derived from the KCP. This parameter can be defined as the difference between the peak value $C(t=t_\text{peak})$ and the late time average (plateau) value $C(t\rightarrow\infty)$ of Krylov complexity, 
\begin{align}\label{eq:diff_peak}
    \Delta C := C(t=t_{\text{peak}}) - C(t\rightarrow\infty) \,.
\end{align}
By definition, $\Delta C \neq 0$ identifies a quantum chaotic system, while $\Delta C = 0$ indicates an integrable one. 

In the bottom panel of Fig.~\ref{Fig:Spread}, we present the KCP order parameter $\Delta C$ as a function of $\kappa$ for various values of the inverse temperature $\beta$. The case at infinite temperature ($\beta=0$), depicted in \color{red}\textbf{red} \color{black}, shows a smooth transition from $\Delta C \approx 0.1$ at $\kappa \rightarrow 0$ to a complete vanishing of $\Delta C$ at large $\kappa$. The critical value of $\kappa$ at which this transition occurs aligns with the critical value $\kappa_c \approx 66$ reported in \cite{Garcia-Garcia:2017bkg} based on spectral statistics methods.
\begin{figure}[t!]
 \centering
     {\includegraphics[width=\linewidth,trim={0 0.05cm 0 0},clip]{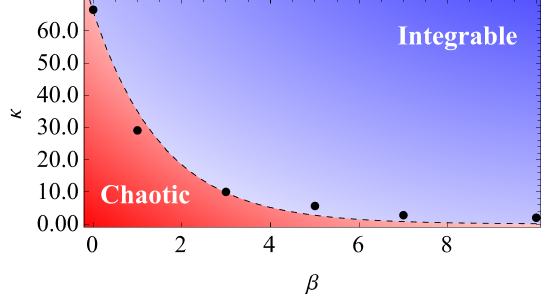} }
\vspace{-0.35cm}
\caption{Phase diagram of the mass-deformed SYK model as a function of the parameter $\kappa$ and the inverse temperature $\beta$. Black dots indicate the critical values of $\kappa$ at which the KCP order parameter, as shown in Fig.\ref{Fig:Spread}, vanishes. The dashed line represents the exponential fit given by Eq.\eqref{fitfit}. The lower region (\color{red}\textbf{red}\color{black}) denotes the chaotic phase, while the upper region (\color{blue}\textbf{blue}\color{black}) represents the integrable phase, with the dashed line marking the boundary between these two phases.
} \label{Fig:Phase}
\end{figure}

As $\beta$ increases and we move away from the infinite temperature limit, three key phenomena are observed. (I) In the quantum chaotic phase (small $\kappa$), the value of $\Delta C$ decreases, indicating that the height of the peak in $C(t)$ relative to the late-time plateau value becomes temperature-dependent. This behavior is consistent with the results for $\kappa = 0$ reported in~\cite{Balasubramanian:2022tpr}. (II) The point of continuous transition from $\Delta C \neq 0$ to $\Delta C = 0$ shifts to lower values of $\kappa$. (III) The width of this transition (e.g., the width is from $\kappa\approx 5$ to $\kappa\approx 30$ for $\beta=0$) increases, with the KCP order parameter exhibiting a smooth crossover rather than a sharp critical transition. 

We define the critical point by identifying the value $\kappa_c$ at which the KCP order parameter vanishes, analogous to the vanishing of the Lyapunov exponent observed in the OTOC analysis of~\cite{Garcia-Garcia:2017bkg}. Using this definition, we construct a phase diagram for the mass-deformed SYK model as a function of the mass deformation parameter $\kappa$ and the inverse temperature $\beta$. This phase diagram is illustrated in Fig.~\ref{Fig:Phase}, with chaotic and integrable phases depicted in {\color{red}\textbf{red}\color{black}} and {\color{blue}\textbf{blue}\color{black}} colors, respectively.

As the temperature decreases, the integrable phase becomes more favorable, and the critical point shifts to smaller values of $\kappa$. We observe that the critical line separating the chaotic and integrable phases is well described by an empirical exponential function:
\begin{align}\label{fitfit}
    \kappa_c (\beta) \approx \kappa_c(0)\, e^{-\frac{2}{\pi} \beta}\,,
\end{align}
which is shown as a black dashed line in Fig.~\ref{Fig:Phase}. This indicates a strong dependence on the inverse temperature $\beta$. This behavior can be potentially rationalized by noting that higher temperatures enable the system to explore a broader range of energy states, leading to a more comprehensive representation of its spectral statistics. Additionally, the observed trend of the critical point with respect to $\beta$ aligns well with previous results obtained using alternative methods~\cite{Garcia-Garcia:2017bkg}.

Having established that the KCP is a signature of quantum chaotic states and a useful order parameter for detecting transitions from chaotic to integrable phases in many-body quantum systems, we now delve deeper into the relationship between Krylov complexity and the spectral form factor (SFF), as defined in Eq.\eqref{eq:SFF}. The mass-deformed SYK model is especially advantageous for this investigation, as it enables a comparative analysis both within the quantum chaotic regime and across the transition to its integrable phase.
\begin{figure}[t!]
 \centering
     {\includegraphics[width=\linewidth,trim={0 0.1cm 0 0},clip]{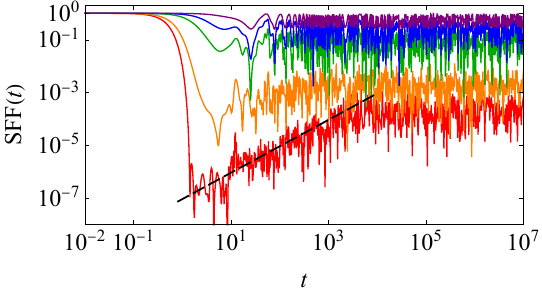} }
     \vspace{-0.3cm}
     \caption{Spectral form factor, Eq.~\eqref{eq:SFF}, as a function of time for $\kappa=1$, $N=26$ and $\beta=0,\,1,\,3,\,5,\,10$ (\color{red}\textbf{red}\color{black}, \color{orange}\textbf{orange}\color{black}, \color{mygreen}\textbf{green}\color{black}, \color{blue}\textbf{blue}\color{black}, \color{darkviolet}\textbf{purple}\color{black}). The black dashed line indicates the average linear growth characteristic of the SFF in the ramp regime.
     } \label{SFFfig}
\end{figure}

A key feature of the SFF for chaotic systems is the presence of an extended linear ramp~\cite{Guhr:1997ve} that follows after the dip. In Fig.~\ref{SFFfig}, we display the time-dependent SFF for various inverse temperatures $\beta$. For the quantum chaotic case with $\beta=0$ (red), the linear ramp is clearly observed as indicated by the black dashed line in Fig.~\ref{SFFfig}. Moreover, we have verified explicitly that the relation with the late time value of Krylov complexity, Eq.~\eqref{eq:saturation}, is obeyed to a high degree of accuracy. As $\beta$ increases, several notable changes occur. First, the $t \rightarrow \infty$ value of SFF$(t)$ rises, approaching the initial value of $1$ as $\beta$ becomes large. Second, the intermediate-time dip in SFF$(t)$ diminishes with increasing $\beta$ and eventually disappears in the large $\beta$ limit. Most significantly, the extent of the linear ramp decreases as $\beta$ increases, ultimately vanishing as the inverse temperature approaches higher values.

Using the empirical function from Eq.\eqref{fitfit} and taking $\kappa_c(0) \approx 66$, we estimate that for $\kappa=1$, the critical value of $\beta$ separating chaotic and integrable phases is approximately $\beta_c \approx 6.58$. This value lies in between the \color{blue}\textbf{blue} \color{black} and \color{darkviolet}\textbf{purple} \color{black} lines in Fig.~\ref{SFFfig}. Our numerical results show that this prediction, derived from the new KCP order parameter, matches the observed disappearance of the ramp in SFF$(t)$. This agreement not only verifies that the presence of a ramp and corresponding dip are clear indicators of quantum chaotic systems, which disappear as the system becomes integrable, but it also independently supports the validity and utility of the order parameter $\Delta C$ as a reliable benchmark for assessing quantum chaotic behavior.

These observations hint at a deeper relationship between Krylov complexity $C(t)$ and the SFF$(t)$ that extends beyond the constraints described in Eq.~\eqref{eq:saturation}. As noted in \cite{Erdmenger_2023}, a potential connection between the KCP and the ramp observed in the SFF seems to be emerging. Additionally, variations of the SFF explored in previous studies \cite{Garcia-Garcia:2017bkg,Nosaka2018} merit further investigation, particularly regarding their relation and potential interplay with Krylov complexity.

Together with the disappearance of the \textit{ramp}, there is an additional observation to be made. As shown in Fig. \ref{Fig:Spread} (bottom), for a fixed value of $\kappa$ and varying $\beta$, there is a critical value at which the KCP disappears. For example, for $\kappa=1$ the KCP disappears in between $\beta=1$ (orange) and $\beta=3$ (green). This transition is also evident in Fig. \ref{SFFfig}, where the dip in the SSF vanishes at around the same value of $\beta$ (green curve). It is therefore plausible that the depth of the dip in the SFF may serve as an alternative order parameter to detect the chaotic to integrable transition, similar to the KCP proposed above. However, as analyzed in detail in Appendix \ref{lala}, this is not the case in general. In fact, we find that this feature is evident only at finite $\beta$, but it disappears in the limit of $\beta \rightarrow 0$. This suggests that the KCP is a more robust indicator of the chaos-integrable transition.

%
\subsection{Sparse SYK model}
To further illustrate the robustness of the KCP as an order parameter for chaotic-integrable transitions, we investigate the sparse SYK model. As with our study of the mass-deformed SYK model, we use a system size of $N=26$ and examine Krylov complexity across different levels of sparsity, parameterized by $p$, or $k$ in Eq.~\eqref{k_order}.

\begin{figure}[t!]
 \centering
     {\includegraphics[width=\linewidth,trim={0 0.1cm 0 0},clip]{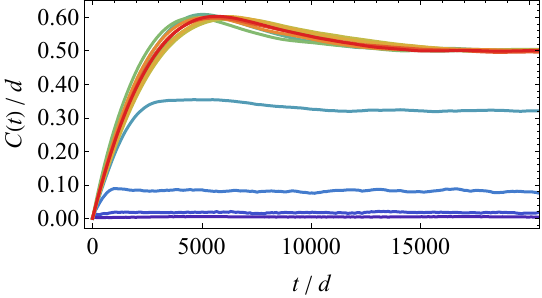} }
     
     {\includegraphics[width=\linewidth,trim={0 0.1cm 0 0},clip]{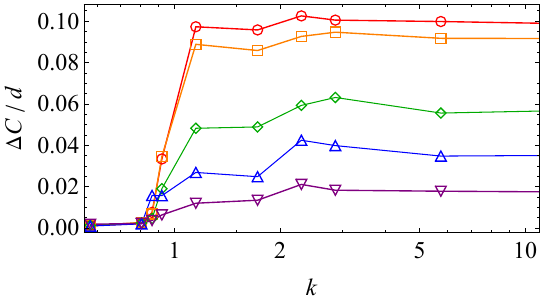} }\vspace{-0.3cm}
\caption{\textbf{Top:} Normalized Krylov complexity $C(t)/d$ with the TFD as initial state for various values of the parameter $p = 1\, (\color{red}\textbf{red} \color{black})$ to $0.001\, (\color{darkviolet}\textbf{purple} \color{black})$ for $N=26$ and $\beta=0$. \textbf{Bottom:} Normalized difference between the peak value $C(t=t_{\text{peak}})$ and the late time value $C(t\rightarrow\infty)$, as a function of $k$ for $\beta=0,\,1,\,3,\,5,\,10$ (\color{red}\textbf{red}\color{black}, \color{orange}\textbf{orange}\color{black}, \color{mygreen}\textbf{green}\color{black}, \color{blue}\textbf{blue}\color{black}, \color{darkviolet}\textbf{purple}\color{black}).} \label{fig:4}
\end{figure}

The top panel of Fig.~\ref{fig:4} illustrates the time evolution of normalized Krylov complexity for various sparsity values in the range $p \in [0.001,1]$ at $\beta=0$. As sparsity increases (\textit{i.e.}, as $p$ decreases), two key observations emerge: (i) the saturation value deviates from $C(t=\infty) = {(d-1)}/{2}$, and (ii) the peak height diminishes.

The deviation in the saturation value arises because, in the sparse SYK model at sufficiently small $p$ (specifically, when $k<1$), a significant number of emergent discrete symmetries, including chiral symmetries, can lead to exact degeneracies in the spectrum~\cite{Garc_a_Garc_a_2021}. These degeneracies modify the late-time behavior of Krylov complexity of the TFD state, resulting in a suppressed saturation value~\cite{Erdmenger:2023wjg,Camargo:2024deu}. This behavior is consistent with previous analyses of Krylov complexity in the sparse SYK model using a different initial state~\cite{Jha:2024nbl}.

\begin{figure}[t!]
 \centering
     {\includegraphics[width=\linewidth,trim={0 0.1cm 0 0},clip]{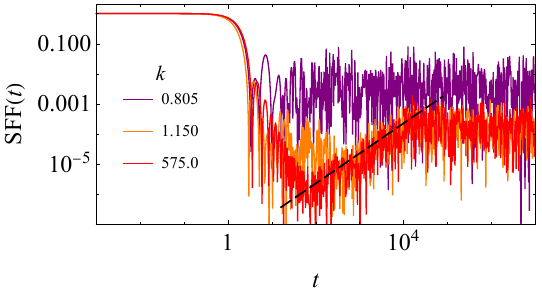} }
     
     {\includegraphics[width=\linewidth,trim={0 0.1cm 0 0},clip]{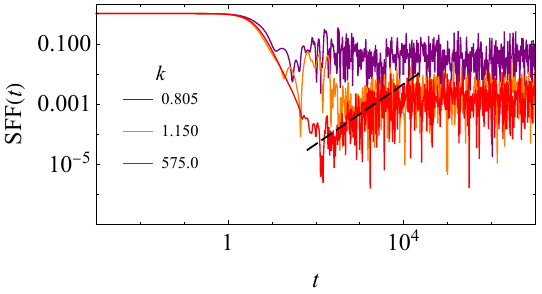} }\vspace{-0.3cm}
\caption{Spectral form factor, Eq.~\eqref{eq:SFF}, as a function of time for various values of the parameter $k = 575\, (\color{red}\textbf{red} \color{black})$, $1.150\, (\color{orange}\textbf{orange} \color{black})$,  $0.805\, (\color{darkviolet}\textbf{purple}\color{black})$ for $N=26$. The black dashed line indicates the average linear growth characteristic of the SFF in the ramp regime.
\textbf{Top:} $\beta=0$. \textbf{Bottom:} $\beta=5$.} \label{fig:5}
\end{figure}

In the bottom panel of Fig.~\ref{fig:4}, we plot the KCP order parameter $\Delta C$ as a function of $k$ for various values of $\beta$. We observe a transition to $\Delta C\approx 0$ with the critical value of $k_c\approx1$ which is consistent with previous studies using spectral statistics methods such as the $r$-parameter statistics~\cite{Garcia-Garcia:2021aa}. Furthermore, Fig.~\ref{fig:5} corroborates this transition, demonstrating that the ramp in the spectral form factor also disappears around $k_c\approx1$.

Our results from both the mass-deformed and sparse SYK models indicate that the KCP serves as an effective order parameter for chaotic-integrable transitions, consistent with findings from conventional spectral statistics methods.
Notably, unlike the mass-deformed SYK model, the critical $k_c$ remains approximately constant at $k_c\approx1$, regardless of the value of $\beta$. This observation may be linked to the level statistics analysis in \cite{Garc_a_Garc_a_2021}, which demonstrated that at $k\approx1$, the system reaches a maximum level of sparsity, leading to the absence of level repulsion.\footnote{For further details, see \cite{Garc_a_Garc_a_2021}, which elaborates on the observation that the distribution width exceeds the level spacing, with the distributions of the first ten eigenvalues being nearly identical.} This suggests that in the regime $k<1$, the Hamiltonian is too sparse to maintain quantum chaotic features. In other words, the Hamiltonian loses its chaotic nature, rendering the influence of the probe (such as $\beta$ through the TFD state) negligible.

\section{Discussion\label{sec:5}} 
In this paper, we proposed that the KCP may serve as a defining feature of quantum chaos and could act as a robust `order parameter' for identifying quantum chaotic phases in many-body systems. We observed that the KCP vanishes in integrable systems while displaying critical dynamics across chaotic-to-integrable quantum and thermal transitions. By computing the KCP using the TFD as an initial state, we effectively identified the chaotic-integrable transitions in the mass-deformed SYK and sparse SYK models at both infinite and finite temperature. This finding is consistent with the results obtained from traditional probes such as spectral statistics and out-of-time-order correlators, and aligns with the independent prediction from the SFF.

Our results provide a compelling answer to the question of which features of Krylov complexity are unique to quantum chaotic systems and how these features change as the system transitions towards integrable regimes. It is now essential to further test our proposal and determine whether the KCP is a universal probe of quantum chaos, comparable to well-established concepts like level repulsion or the quantum Lyapunov exponent. In this vein, it is important to gain a deeper understanding of potential counterexamples, such as quantum systems with integrable phases exhibiting saddle-dominated scrambling, like the Lipkin-Meshkov-Glick model \cite{Bhattacharjee:2022vlt,Huh:2023jxt},  or quantum systems with a mixed phase space, such as the stringy matrix models recently considered in \cite{Amore:2024ihm}.

Lastly, our analysis underscores the importance of the TFD state in the study of quantum chaos in many-body systems, building on previous observations~\cite{Camargo:2024deu}. This state is crucial for Krylov complexity to effectively probe random matrix physics, likely because random matrix behavior encompasses the entire level-spacing spectrum of the system. Consequently, any measure aimed at probing random matrix behavior benefits from the TFD state's ability to encompass the full spectrum. TFD states also play a significant role in holography, serving as CFT duals of two-sided black holes~\cite{Maldacena_2003}. Recent findings suggest a connection between the Krylov complexity of chord states in the double-scaled SYK model and the length of the dual Lorentzian wormhole in Jackiw-Teitelboim gravity~\cite{Rabinovici:2023yex}, a lower-dimensional realization of holography. Therefore, our work might provide valuable insights into the holographic dual description of chaotic-integrable transitions in quantum systems and contribute to addressing the profound and long-standing `million-dollar question' of quantum gravity.

Furthermore, it may be worthwhile to explore the connection between Krylov complexity and more classical approaches to integrability. One interesting direction could involve an ``extended" definition of quantum chaos inspired by the Liouville-Arnold theorem, stating that a system with $N$ degrees of freedom is considered \textit{regular} if the number of the first integrals, $M$, equals $N$, while the system is deemed \textit{chaotic} when $M$ is less than $N$.

%
\acknowledgments
We would like to thank {Antonio M. Garcia-Garcia} for collaboration at the initial stage of this project and {Junggi Yoon} for valuable suggestions. We would like to thank {Pratik Nandy and Debodirna Ghosh} for useful comments on a first version of this manuscript. MB would also like to thank Dario Rosa for illuminating discussions on complexity and quantum chaos.  
MB and KBH acknowledge the support of the Shanghai Municipal Science and Technology Major Project (Grant No.2019SHZDZX01). M.B. acknowledges the sponsorship from the Yangyang Development Fund.
HSJ and JFP are supported by the Spanish MINECO ‘Centro de Excelencia Severo Ochoa' program under grant SEV-2012-0249, the Comunidad de Madrid ‘Atracci\'on de Talento’ program (ATCAM) grant 2020-T1/TIC-20495, the Spanish Research Agency via grants CEX2020-001007-S and PID2021-123017NB-I00, funded by MCIN/AEI/10.13039/501100011033, and ERDF A way of making Europe.
KYK was supported by the Basic Science Research Program through the National Research Foundation of Korea (NRF) funded by the Ministry of Science, ICT $\&$ Future Planning (NRF-2021R1A2C1006791) and the Al-based GIST Research Scientist Project grant funded by the GIST in 2024. KYK was also supported by the Creation of the Quantum Information Science R$\&$D Ecosystem (Grant No. 2022M3H3A106307411) through the National Research Foundation of Korea (NRF) funded by the Korean government (Ministry of Science and ICT).
All authors contributed equally to this paper and should be considered as co-first authors. 

%
\appendix
\section{Details on the numerical methods\label{sec:App}}
In this appendix, we provide additional details on the numerical computations discussed in the main text, along with further analyses to substantiate our results. Here, we primarily focus on the mass-deformed SYK model; however, the key features are consistent with those observed in the sparse SYK model as well.

\subsection{Block diagonalization of the SYK Hamiltonian}
Block diagonalization of the Hamiltonian matrix is crucial for analyzing spectral statistics of the energy spectrum. By decomposing the Hamiltonian into smaller blocks based on the system's symmetries, the statistical properties of the energy levels can be studied more efficiently. This approach simplifies computations, improves numerical accuracy, and reduces the effective size of the system.

\begin{figure}[h!]
 \centering
     {\includegraphics[width=\linewidth,trim={0 0 0 0},clip]{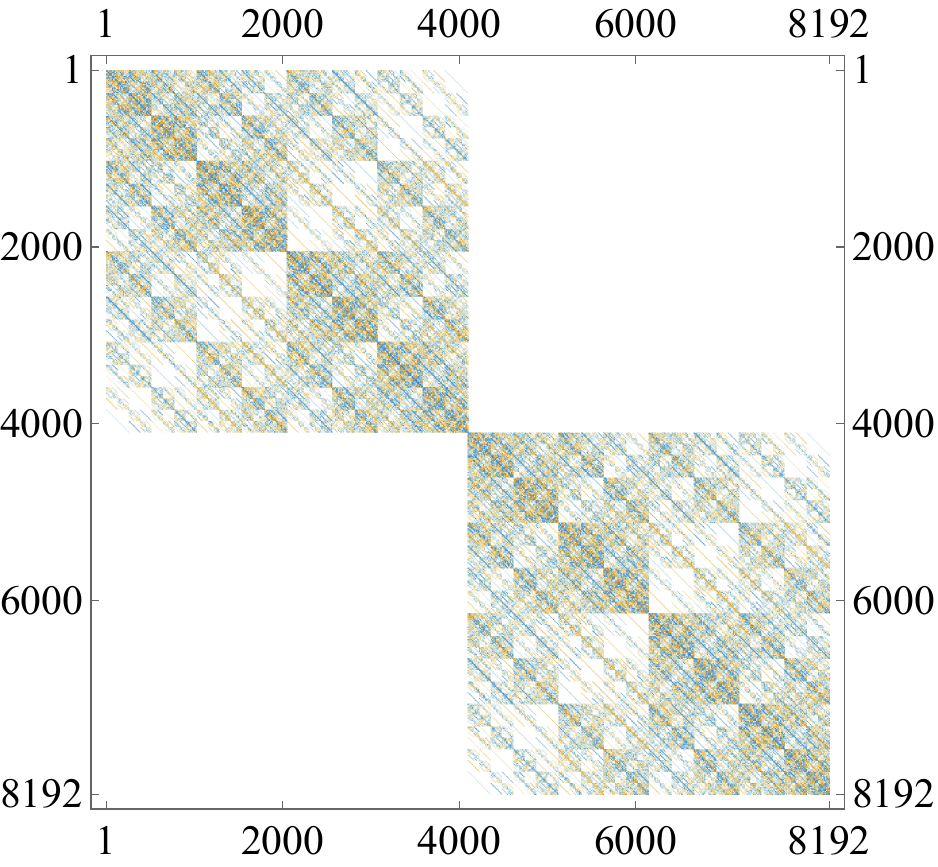} }
     \vspace{-0.3cm}
     \caption{The block diagonalized mass-deformed SYK model for $N=26$ and $\kappa=1$. The top-left block corresponds to the parity-odd sector, while the bottom-right block corresponds to the parity-even sector.} \label{Fig:Matp}
\end{figure}

Regarding the SYK model, its Hamiltonian possesses a conserved charge parity operator $P$~\cite{You_2017,Cotler_2017,Krishnan_2018},
\begin{align}
    P =  \left\{ \begin{array}{ll}
         (-i \chi_1 \chi_2)(-i \chi_3 \chi_4)\,\cdots\,(-i \chi_{N-1} \chi_{N} )   & \quad N \in \text{even}\,, \cr
         (-i \chi_1 \chi_2)(-i \chi_3 \chi_4)\,\cdots\,(-i \chi_{N} \chi_\infty)   & \quad N \in \text{odd} \,,
        \end{array}\right.
\end{align}
where $\chi_i\,(i=1,2, \dots , N)$ are Majorana fermions, and $N$ is the system size. The SYK Hamiltonian $H$ commutes with the parity operator $P$, $[H, P] = 0$, allowing the Hamiltonian to be block diagonalized. Using an invertible matrix consisting of the eigenvectors of $P$, the system is split into parity-even and parity-odd sectors, each with a dimension $d = 2^{\frac{N}{2}-1}$. Fig.~\ref{Fig:Matp} illustrates the typical block diagonalized Hamiltonian of a mass-deformed SYK model at finite $\kappa$.

In our study, we have focused on numerical computations within the parity-even sector for $N=26$, using both \textsc{Wolfram Mathematica} and \textsc{Matlab} to cross-verify our results. We have also confirmed that the parity-odd sector produces similar outcomes for Krylov complexity.

%
\subsection{Lanczos coefficients and quantum chaos}
In the main text, we focused on the Krylov complexity derived from solving the Schrödinger equation using the given Lanczos coefficients $\{a_n,\,b_n\}$. Here, we examine the Lanczos coefficients $\{a_n,\,b_n\}$ of the mass-deformed SYK models. As suggested in references~\cite{Hashimoto:2023swv,Parlett_1998}, the Lanczos coefficients can be obtained through the Lanczos algorithm, which minimizes numerical errors in the orthogonalization process, ensuring consistency with those derived from the Hessenberg form. The algorithm is outlined as follows:
\begin{enumerate}
\item Initialize with $b_0 := 0$, $| K_0 \rangle := | \psi(0) \rangle$, $a_0 := \langle K_0 | \mathcal{D} | K_0 \rangle$, where $\mathcal{D} := \text{diag}\left(E_1, \ldots, E_{d}\right)$.
\item For $n \geq 1$: Compute $| A_n \rangle = \left( \mathcal{D} - a_{n-1} \right) | K_{n-1} \rangle - b_{n-1} | K_{n-2} \rangle$.
\item Replace $|A_n\rangle \rightarrow |A_n\rangle - \sum_{m=0}^{n-1} \langle{A_m}|{K_0}\rangle  |A_m\rangle$.
\item Set $b_n = \langle{A_n}|{A_n}\rangle^{1/2}$.
\item If $b_n = 0$, terminate the algorithm; otherwise, set $|{K_n}\rangle = b_n^{-1} |{A_n}\rangle$ and $a_n = \langle{K_n}| \mathcal{D} |{K_n}\rangle$, then return to step 2.
\end{enumerate}

In our study, we utilize the TFD state as the initial state $| \psi(0) \rangle$. To illustrate our numerical results, we present specific computations for the maximally entangled state, i.e., the TFD state with $\beta = 0$. Using the Lanczos algorithm, we obtain the Lanczos coefficients for the mass-deformed SYK model, as shown in Fig.~\ref{SMFIG1} below.
We observe that $a_n$ exhibits oscillatory behavior, while $b_n$ initially grows (see inset), reaches a peak, and then diminishes as $n$ approaches the dimension of the Krylov space. These characteristics are consistent with those observed in the original SYK model ($\kappa=0$) \cite{Balasubramanian:2022tpr} and in RMT \cite{Erdmenger:2023wjg}.
\begin{figure}[h!]
\centering
{\includegraphics[width=\linewidth]{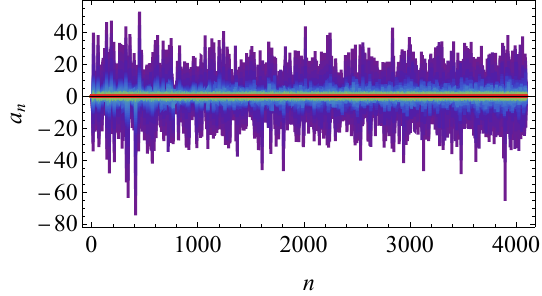}}
\hfill
{\includegraphics[width=\linewidth]{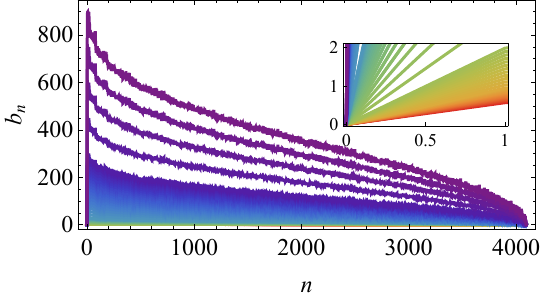}}
\vspace{-0.3cm}
\caption{Lanczos coefficients $\{a_n,\,b_n\}$ for the mass-deformed SYK model for various values of the mass-deformation parameter, ranging from $\kappa = 0\, (\color{red}\textbf{red} \color{black})$ to $\kappa =300\, (\color{darkviolet}\textbf{purple} \color{black})$.} \label{SMFIG1}
\end{figure}

\begin{figure}[h!]
\centering
{\includegraphics[width=\linewidth]{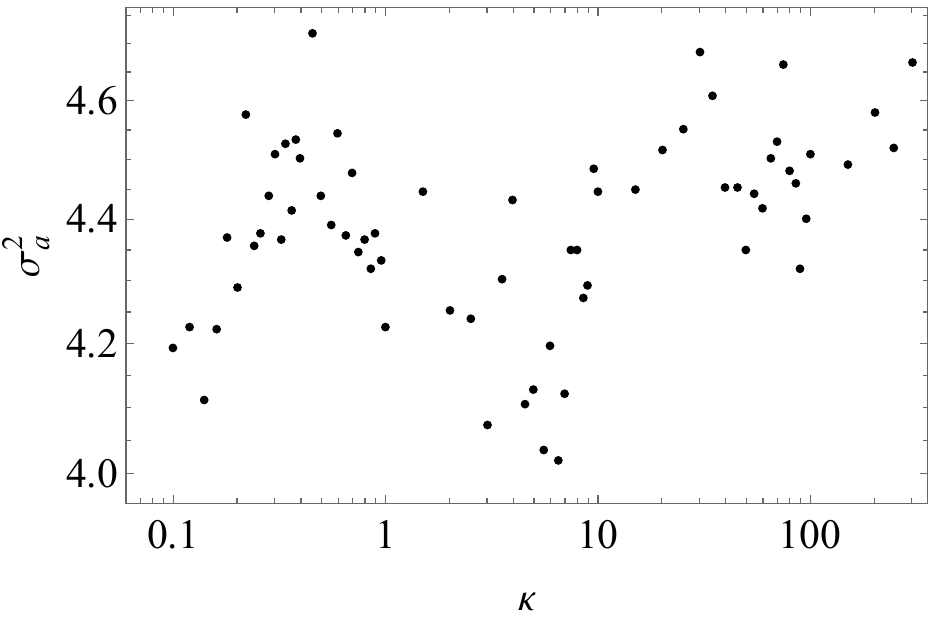}}
\qquad
{\includegraphics[width=\linewidth]{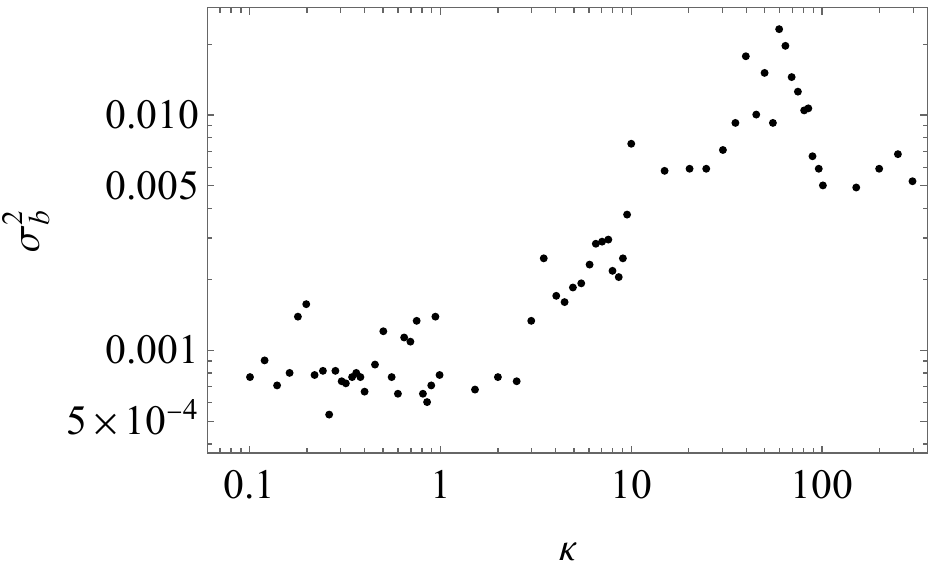}}
\vspace{-0.3cm}
\caption{The variance of the Lanczos coefficients $\sigma_{a, b}^2$ as a function of $\kappa$.} \label{SMFIG2}
\end{figure}

\begin{figure}[]
\centering
{\includegraphics[width=\linewidth]{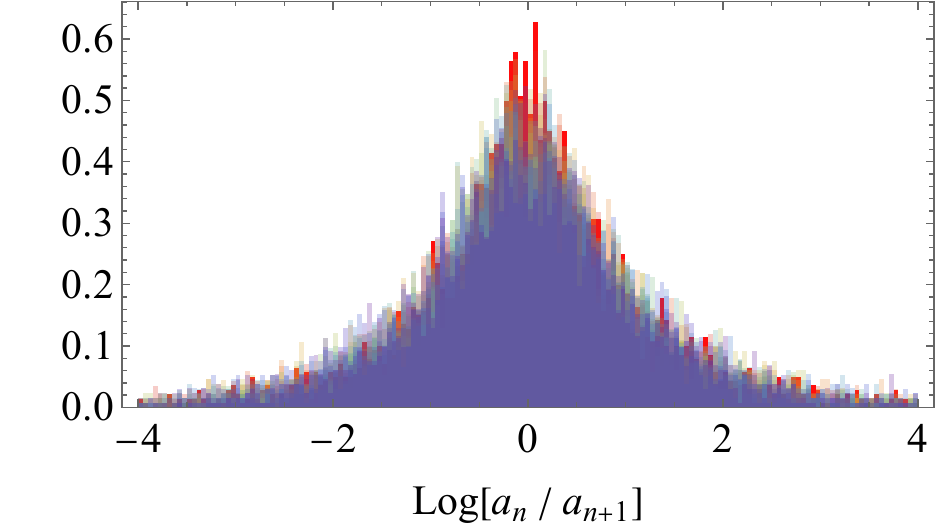}}
\hfill
{\includegraphics[width=\linewidth]{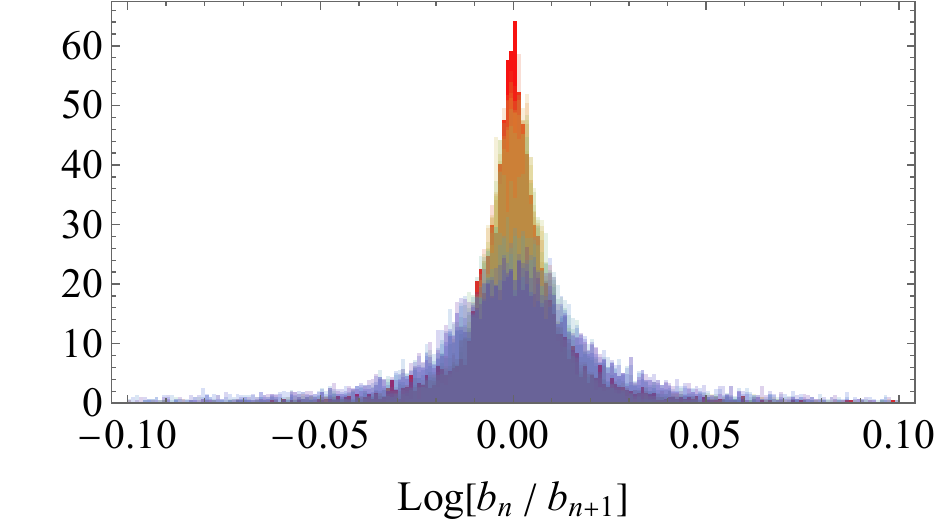}}
\vspace{-0.3cm}
\caption{Histogram of the Lanczos coefficients $a_n$ (left panel) and $b_n$ (right panel) for values of $\kappa$ ranging from $\kappa = 0\, (\color{red}\textbf{red} \color{black})$ to $\kappa = 300\, (\color{darkviolet}\textbf{purple} \color{black})$.} \label{SMFIG3}
\end{figure}

Two notable observations regarding the effect of $\kappa$ are as follows. First, as $\kappa$ increases, the slope of the initial growth of $b_n$ also increases (see the inset), indicating that the value of $b_1$ is enhanced by the mass deformation parameter. This enhancement may explain the behavior of the slope of $C(t)$ in Fig.~\ref{Fig:Spread} with increasing $\kappa$, as the early-time behavior of the Krylov complexity has been shown to follow $C(t\ll1)\approx b_1^2 t^2$~\cite{Huh:2023jxt}. The second observation concerns the variance of the Lanczos coefficients, defined as~\cite{Hashimoto:2023swv}
\begin{align}\label{}
\begin{split}
    \sigma_{a}^2 &:= \text{Var} \left(x_i^{(a)}\right) \,, \quad x_i^{(a)} := \log \left( \frac{a_{2i-1}}{a_{2i}} \right) \,, \\
    \sigma_{b}^2 &:= \text{Var} \left(x_i^{(b)}\right) \,, \quad x_i^{(b)} := \log \left( \frac{b_{2i-1}}{b_{2i}} \right) \,.
\end{split}
\end{align}
For additional information on the variance of Lanczos coefficients in the context of Krylov operator complexity, see~\cite{Rabinovici:2021qqt,Rabinovici:2022beu}.
We observe that the variance increases in the integrable regime (large $\kappa$) compared to the chaotic regime (small $\kappa$). This trend is especially pronounced for $\sigma_{b}^2$; see Fig.~\ref{SMFIG2}. 
However, unlike the KCP, we do not observe a distinct feature indicative of a critical phase transition around $\kappa_c \approx 66$. This suggests that the KCP may serve as a more effective order parameter than the Lanczos coefficients themselves.
Refer to Fig.~\ref{SMFIG3} for the histogram plot of the Lanczos coefficients' distribution, which supports the conclusions drawn from the variance analysis.
%

%
\subsection{Normalization condition and Ehrenfest theorem}
We have confirmed that our numerical results meet essential consistency checks, including the wave function normalization condition~\cite{Balasubramanian:2022tpr} and the Ehrenfest theorem~\cite{Erdmenger_2023}.

\begin{figure}[h!]
\centering
{\includegraphics[width=\linewidth]{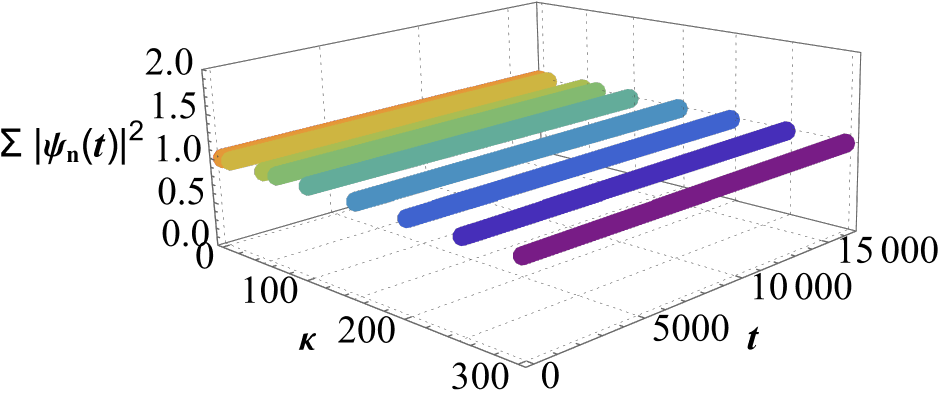}}
\vspace{-0.3cm}
\caption{The normalization of the Krylov wave functions.} \label{SMFIG4}
\end{figure}

\begin{figure*}[]
\centering
\begin{minipage}{0.29\textwidth}
        \centering
        \includegraphics[width=\textwidth]{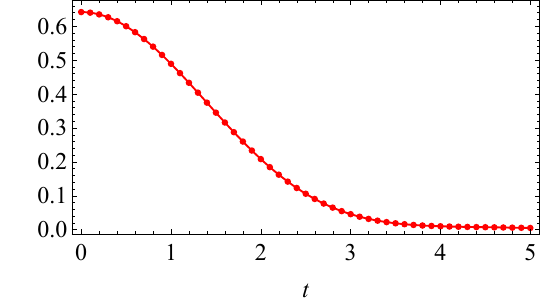}
\end{minipage}
\quad
\begin{minipage}{0.29\textwidth}
        \centering
        \includegraphics[width=\textwidth]{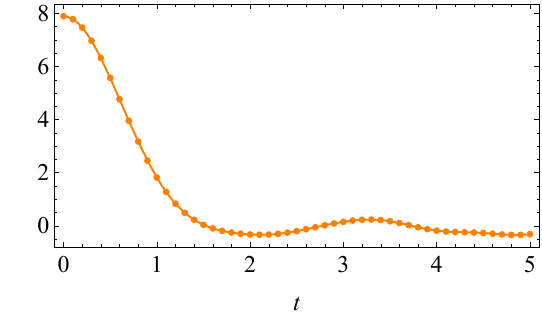}
\end{minipage}
\quad
\begin{minipage}{0.29\textwidth}
        \centering
        \includegraphics[width=\textwidth]{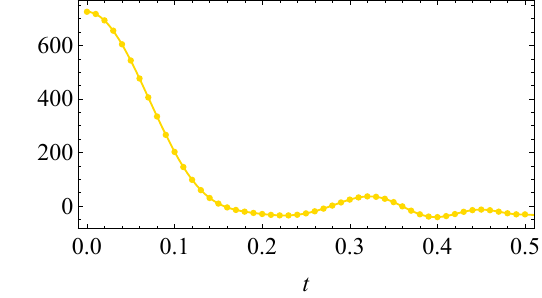}
\end{minipage}

\begin{minipage}{0.31\textwidth}
        \centering
        \includegraphics[width=\textwidth]{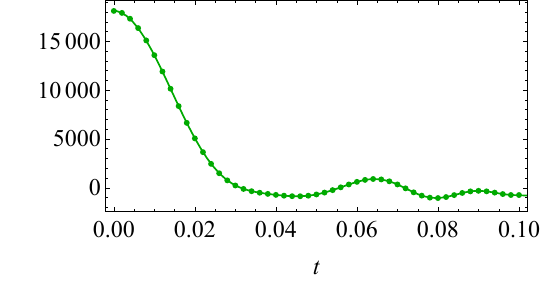}
\end{minipage}
\quad
\begin{minipage}{0.31\textwidth}
        \centering
        \includegraphics[width=\textwidth]{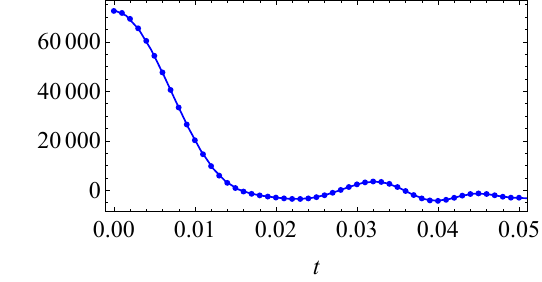}
\end{minipage}
\quad
\begin{minipage}{0.31\textwidth}
        \centering
        \includegraphics[width=\textwidth]{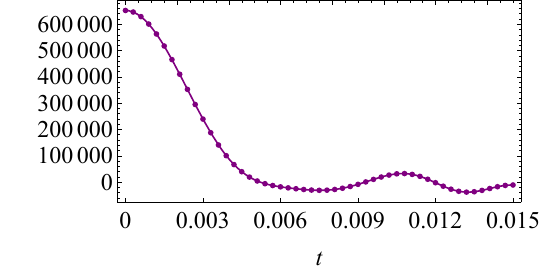}
\end{minipage}
\caption{Ehrenfest theorem in the mass-deformed SYK models for $\kappa=0,\,1,\,10,\,50,\,100,\,300$ (\color{red}\textbf{red}\color{black}, \color{orange}\textbf{orange}\color{black}, 
\color{yellow}\textbf{yellow}\color{black}, 
\color{mygreen}\textbf{green}\color{black}, \color{blue}\textbf{blue}\color{black}\,\,and \color{darkviolet}\textbf{purple}\color{black}, respectively). The solid lines correspond to the L.H.S. of \eqref{EHRTH2}, while the dots represent the R.H.S. of \eqref{EHRTH2}.}\label{SMFIG5}
\end{figure*}

First, we address the normalization condition for the Krylov wave functions,
\begin{align}\label{}
  \sum_{n} |\psi_n(t)|^2 = 1 \,,
\end{align}
which ensures the unitarity of time evolution.
As shown in Fig.~\ref{SMFIG4}, this normalization condition is satisfied within our time window for numerical computations of Krylov complexity, specifically for $t \leq 2^{N/2+1}$. 
To maintain normalization over a longer time window, including the late-time regime, it is necessary to consider larger values of $n_{\text{max}}$, as discussed in~\cite{Hashimoto:2023swv,Camargo:2023eev}. Our results indicate that the chosen cutoff value of $n_{\text{max}}=d$ is adequate for capturing all key features of Krylov complexity ---the initial ramp, peak, decline, and plateau--- while satisfying the normalization condition.
Second, an important aspect of Krylov complexity is its adherence to the Ehrenfest theorem~\cite{Erdmenger:2023wjg}, expressed as:
\begin{align}\label{EHRTH}
\begin{split}
\partial^2_t \langle \psi | \hat{C} | \psi \rangle = - \langle \psi |  \left[ \left[ \hat{C}\,, \mathcal{L} \right], \mathcal{L} \right] | \psi \rangle\,
\end{split}
\end{align}
where $\hat{C}:= \sum_n n\,|K_n \rangle \langle K_n |$ 
and $\mathcal{L} = H \otimes \mathbb{I}$ denotes the Liouvillian and $\mathbb{I}$ represents the identity operator. By applying the Schrödinger equation along with the definition of Krylov complexity, the Ehrenfest theorem \eqref{EHRTH} can be formulated in terms of the Lanczos coefficients and the Krylov wave functions as follows:
\begin{align}\label{EHRTH2}
\begin{split}
\partial^2_t C(t) &= 2 \sum_n  \left[ \left( b_{n+1}^2 - b_n^2 \right) \psi_n(t) \psi_n^{*}(t) \right. \\ &\left. +\left(a_{n+1} - a_{n}\right)b_{n+1}  \psi_{(n+1}(t) \psi_{n)}^{*}(t)  \right]\,,
\end{split}
\end{align}
where $\mathcal{T}_{(a} \tilde{\mathcal{T}}_{b)}:=\frac{1}{2}\left(\mathcal{T}_{a}\tilde{\mathcal{T}}_{b}+\mathcal{T}_{b}\tilde{\mathcal{T}}_{a}\right)$.
In Fig.~\ref{SMFIG5}, we validate the Ehrenfest theorem for mass-deformed SYK models, demonstrating the relationship between the second time derivative of Krylov complexity and a combination of Lanczos coefficients. It is important to emphasize that this relationship, as expressed in \eqref{EHRTH2}, holds universally for any system by construction. This verification strengthens the reliability of our numerical results.
%

%

%
\begin{figure}[h!]
\centering
{\includegraphics[width=\linewidth]{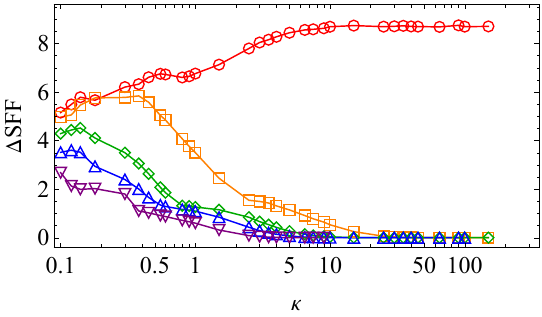}}
\qquad
{\includegraphics[width=\linewidth]{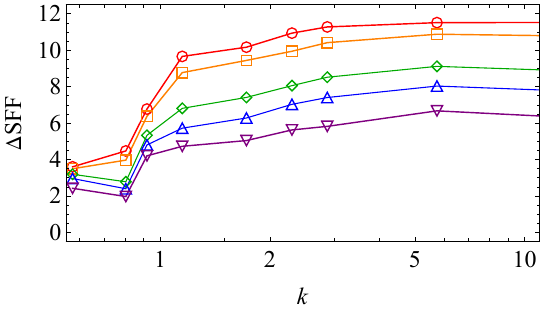}}
\vspace{-0.3cm}
\caption{$\Delta\text{SFF}$, Eq.~\eqref{DSFF}, as a function of  parameter $\kappa$ and $k$ (\textbf{Top:} Mass-deformed SYK model, \textbf{Bottom:} Sparse SYK model) for $\beta=0,\,1,\,3,\,5,\,10$ (\color{red}\textbf{red}\color{black}, \color{orange}\textbf{orange}\color{black}, \color{mygreen}\textbf{green}\color{black}, \color{blue}\textbf{blue}\color{black}, \color{darkviolet}\textbf{purple}\color{black}).} \label{DSFFFIG}
\end{figure}

\subsection{The depth of the SFF-hole}\label{lala}
We further explore the dynamics of the \textit{dip} in the SFF as the system transitions from chaotic to integrable regimes. Specifically, we observe that the depth of the dip tends to be suppressed along this transition.

To quantify this behavior, we utilize Eq.~\eqref{eq:SFF} and define
\begin{align}\label{DSFF}
\begin{split}
\Delta \text{SFF} = \text{Log}\left[\text{SFF}(t_\text{dip})\right] -\text{Log}\left[\text{SFF}(t=\infty)\right]\,,
\end{split}
\end{align}
which characterizes the difference between the dip and the saturation value of the SFF.

Fig.~\ref{DSFFFIG} presents the computed $\Delta \text{SFF}$ for the SYK models discussed in the main text. A comparison with the KCP, as shown in the bottom panels of Figs.~\ref{Fig:Spread} and \ref{fig:4}, reveals a notable similarity in the behavior. However, we find that the KCP serves as a more robust indicator of the chaotic-to-integrable transition. In particular, $\Delta \text{SFF}$ fails to effectively capture the transition in the limit of $\beta=0$ for the mass-deformed SYK model (red data in the top panel of Fig.~\ref{DSFFFIG}). This is not a surprise as the depth of the dip is still manifest for the SYK$_2$ model (corresponding to $\kappa\rightarrow\infty$ case) in the $\beta=0$~\cite{Erdmenger:2023wjg}.

%

%

%
\end{document}